# Direct Observation of *sp-d* Exchange Interaction in Mn$^{2+}$ doped All-inorganic Perovskite Quantum Dots (CsPbX$_3$: X= Cl, Br)

Prasenjit Mandal [b] and Ranjani Viswanatha*,[a, b]

[a] *International Centre for Material Science (ICMS), JNCASR, Bangalore*

[b] *New Chemistry Unit (NCU), JNCASR, Bangalore*

*Corresponding author, email: rv@jncasr.ac.in

**Abstract**

The field of lead halide perovskite nanocrystal doping has witnessed notable progress in recent times, leading to the creation of innovative materials that showcase compelling physical characteristics and hold substantial technological promise. The true characteristics of these materials lie in the presence of dopant-carrier magnetic exchange interactions. This work presents the first direct observation of such exchange interactions in colloidal Mn-doped CsPbX$_3$ (X= Cl, Br) quantum dots (QDs). Here, we employ magnetic circular dichroism (MCD) spectroscopy to unambiguously demonstrate the successful doping and the presence of giant excitonic Zeeman splitting in CsPbX$_3$ (X= Cl, Br) QDs doped with Mn$^{2+}$. The controllable tuning of effective exciton *g*-factors ($g_{eff}$) within the range of 2.1 to (-)314 has been achieved through the process of doping with 6.9 % Mn in CsPbCl$_3$, which will facilitate their application towards future spintronics.

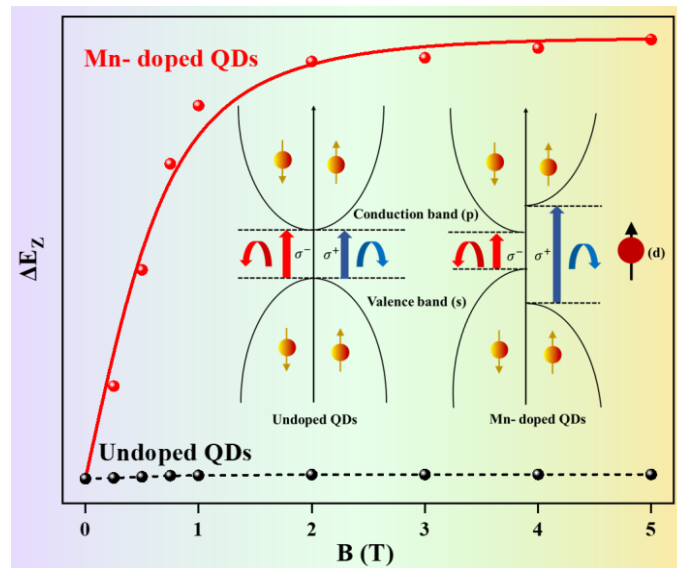

**Figure** TOC of the paper

**Keywords:** Perovskite quantum dots, doping, *sp-d* exchange interaction, magnetic circular dichroism (MCD), etc.

**Introduction**

Diluted magnetic semiconductor (DMS) materials,[1-7] which merge the characteristics of semiconductors and magnetic materials, exhibit great potential as promising candidates for diverse applications such as spintronic devices,[8] magneto-optical devices,[9] quantum information processing,[10] etc. All of these applications stem from the exchange interactions between carriers and dopants within the DMSs, known as *sp-d* exchange interactions.[11] These interactions facilitate the control of carrier spins through the magnetization of magnetic impurities. Thus, the incorporation of external magnetic impurities within host nanocrystals (NCs) plays a critical role in exploring dopant-induced properties. In this regard, lead halide perovskite-based colloidal semiconductor NCs have emerged as promising materials due to their solution processability, tuneable electronic properties, size-dependent optical effects, and facile doping.[12]

Different transition metal ions, such as manganese (Mn), cobalt (Co), nickel (Ni), and iron (Fe), can be incorporated into the crystal lattice of perovskite QDs during their synthesis.[13-16] Among the various transition metal ions, the most commonly utilized is $Mn^{2+}$ ions due to their ability to produce a yellow-orange long-lived emission through spin-forbidden *d-d* transitions. Strong sensitized luminescence observed from the *d–d* transition of $Mn^{2+}$ ions when the $CsPbX_3$ host is excited at the band-edge suggests a significant exchange coupling between the host's charge carriers and the dopant *d* electrons. This coupling plays a vital role in mediating the energy transfer, which is necessary for attaining the unique properties of magnetically doped QDs. Long-lived Mn emission can be coupled vibrationally with the host and give rise to delayed emission at cryogenic temperatures for Mn-doped $CsPbCl_3/CsPb(Cl/Br)_3$. However, such delayed emission can be observed at room temperature in $CsPbBr_3$. Pradeep et al. have shown that second excited states of the host rather than the first excited states, are involved in facilitating the delayed emission through compositional modulation and the confinement effect over various Mn-doped NCs.[17] Thus, the understanding of electronic structure and state-specific coupling mechanism between the host and dopant is important to explain these phenomena.

The *s, p-d* coupling can result exchange interactions between the spin of the free carriers and the localized *d*-electrons, leading to splitting of their band structure known as "giant" Zeeman splitting. The exchange interaction between *s, p-* charge carriers and transition metal ions *d*-electrons in doped II-VI semiconductor QDs has been thoroughly investigated, yielding valuable insights on multiple properties. In this regard, MCD provides a potent means of qualitatively examining the existence of *sp-d* exchange interactions through the direct demonstration of Zeeman splitting.[18-19] This lays the foundation for manipulating magneto-optical properties in order to facilitate technological applications. For instance, Gamelin et al. were the first to report the direct observation of the *s, p-d* exchange interaction and giant Zeeman splitting in QDs using MCD spectroscopy, providing direct evidence of the doping of $Mn^{2+}$ and $Co^{2+}$ ions in CdSe QDs.[20] They also discussed the thermal tunability of this interaction in a separate report.[21] Vlaskin et al. elucidated the impact of shell growth on the magnetic exchange interactions between dopants and carriers in $Zn_{1-x}TM_xSe$ core, $Zn_{1-x}TM_xSe/CdSe$ inverted core/shell, and $Zn_{1-x}TM_xSe/ZnSe$ isocrystalline core/shell structures doped with $Co^{2+}$ and $Mn^{2+}$ using MCD.[22] Recently, couple of work from Gamelin et al. highlight the spin precession of vapour deposited thin films and NCs of $CsPbBr_3$ using MCD and time resolved faraday rotation (TRFR).[23-24] However, the impact of the *sp-d* exchange interaction on perovskite has yet to be thoroughly investigated. Nevertheless, a recent study by Zhang et al.[25] utilized a combination of magnetic field and circular polarized laser excitation to amplify the spin polarization in Mn-doped $CsPbCl_3$. This indicates that the excitonic characteristics of these systems can be bolstered through the utilization of polarized light and a magnetic field.

With the aforementioned knowledge, we include a small quantity of $Mn^{2+}$ in $CsPbX_3$ (X = Cl, Br) in this study, and we vary the halide ion ratio to study its effect on coupling strength. To explore this, MCD spectroscopy is employed to analyse the *sp-d* exchange interactions in these DMS NCs by detecting substantial excitonic Zeeman splitting. This is achieved through differential absorption of circularly polarized light, specifically left ($\sigma^-$) and right ($\sigma^+$) circular polarization in presence of magnetic field. This is the first direct observation of *sp-d* exchange interactions in colloidal $Mn^{2+}$-doped $CsPbX_3$ (X= Cl, Br) QDs. The MCD spectra of Mn-doped $CsPbCl_3$ reveal the direct observation of *sp-d* exchange interaction, with Mn strongly coupled to the higher excited state rather than the band edge, resulting in a giant Zeeman splitting. Effective exciton *g*-factor can be achieved up to ~ (-) 314 by varying the Mn-concentration in $CsPbCl_3$. We also observed that with the variation of composition with the incorporation of Br, the exchange interaction diminishes, consequently weakening the Zeeman splitting. The findings presented in this study demonstrate the promising potential of $Mn^{2+}$-doped $CsPbX_3$ (X= Cl, Br) QDs as materials for investigating spin effects in semiconductor nanostructures.

**Experimental section**

**Chemicals**

The following chemicals were employed in the experiments: $Cs_2CO_3$ (99.9%), Oleylamine (OAm) (70% technical grade), Oleic acid (OAc) (90% technical grade), 1-Octadecene (ODE) (90% technical grade), $PbCl_2$ (99.9%), $PbBr_2$ (98%), $MnCl_2 \cdot 4H_2O$ (99%), $MnBr_2$ (98%) (n-hexane (>97.0%), anhydrous methyl acetate (99.5%), and trioctylphosphine (TOP). All these chemicals were obtained from Sigma Aldrich. Hexane (AR grade) was purchased from Thermo Fischer Scientific India Pvt. Ltd. None of the chemicals underwent purification prior to usage.

**Method**

**Preparation of cesium oleate (Cs-oleate)**

In a 50 mL three-necked round bottom flask, 20 mL of ODE and 1.5 mL of OAc are combined with 400 mg (1.23 mmol) of $Cs_2CO_3$. The flask was degassed, which involves removing moisture from the solution, for 1.25 hours while continuously stirring at 120 °C. This step helps ensure a clean reaction environment. After degassing, argon gas was purged into the degassed solution for 15 to 20 minutes at 120 °C to create an oxygen-free environment and prevent unwanted reactions. Then, the temperature was gradually raised to 140 °C in an argon environment and maintained at that temperature until all of the $Cs_2CO_3$ had reacted with the oleic acid. This reaction produces Cs-oleate, which is the desired product. The solution was cooled to room temperature and transferred into vials filled with argon to prevent oxidation or contamination. These vials are stored for later use. To obtain a clean solution before the hot injection process, the solution was heated to dissolve any precipitate that may have formed during the reaction.

**Synthesis of undoped and Mn-doped NCs**

We utilized the hot injection technique reported by Protesescu et al.[26] to produce a series of Mn-doped perovskite NCs with varying halide compositions. In a typical procedure, a three-neck round bottom flask was used, into which 0.188 mmol of Pb-halide salt, 5 mL of ODE, 1.5 mL of OAc, and 1.5 mL of OAm were added. The flask contents were stirred and degassed for 1 hour at 120°C to remove moisture. Subsequently, a specific amount of TOP was introduced, followed by another 15-minute degassing step. The temperature was then gradually raised to 170°C in an argon atmosphere. At this temperature, 0.4 mL of pre-heated Cs-oleate was rapidly injected into the flask, and the reaction was promptly quenched using an ice bath. A similar synthesis procedure was employed to obtain doped nanocrystals with the stoichiometric mixtures of $PbCl_2$, $PbBr_2$, $MnCl_2 \cdot 4H_2O$, and $MnBr_2$. The details amount of the precursors, along with the sample codes, are listed below in Table T1. The specific concentration of the dopant obtained through ICP-OES is also included.

**Table T1** Details of Precursors used for the Synthesis of Pure and Mixed Halide Doped NCs.

| Sample Name | $PbCl_2$ (mg) | $PbBr_2$ (mg) | $MnCl_2 \cdot 4H_2O$ (mg) | $MnBr_2$ (mg) | TOP (mL) | Actual Mn % obtained from ICP-OES |
|---|---|---|---|---|---|---|
| P1- $CsPbCl_3$ | 52.3 | - | - | - | 1.0 | - |
| Sample Name | $PbCl_2$ (mg) | $PbBr_2$ (mg) | $MnCl_2 \cdot 4H_2O$ (mg) | $MnBr_2$ (mg) | TOP (mL) | Actual Mn % obtained from ICP-OES |
| MnP1 (Mn 2.1%) | 28 | - | 79 | - | 1.0 | 2.1 % |
| MnP1 (Mn 6.9%) | 28 | - | 118.7 | | 1.0 | 6.9 % |
| MnP3 | 31.3 | 27.6 | 89.3 | 64.6 | 0.5 | 2.7 % |
| MnP4 | 20.9 | 41.4 | 59.5 | 96.9 | 0.5 | 2.2 % |
| MnP5 | 10.4 | 55.2 | 29.7 | 129.2 | 0.5 | 1.4 % |
| MnP6 | - | 69 | - | 161.5 | - | 1.5% |

**Purification of synthesized NCs**

Following the ice bath quenching procedure described earlier, the flask's contents underwent centrifugation at a speed of 5000 rpm for a duration of 10 minutes. Subsequently, the supernatant was carefully discarded, and the particles were re-dispersed in 4 mL of hexane.

The resulting mixture was refrigerated for a period of 24 hours. Then, the supernatant was subjected to a wash using methyl acetate, while the precipitate obtained was dissolved in hexane. The resulting solution was kept refrigerated for subsequent characterization.

**Characterization and spectroscopic studies**

Several techniques were employed to characterize and analyse the synthesized NCs. The UV-visible absorption spectra of perovskite NCs dissolved in hexane were recorded using an Agilent 8453 UV-visible spectrometer. The FLSP920 spectrometer by Edinburgh Instruments was utilized for acquiring steady-state PL spectra. A 450 W xenon lamp served as the source. Meanwhile, within the same instrument, the PL lifetime measurements were conducted using the EPL-405 pulsed diode laser and EPL-340 diode laser as the excitation sources ($\lambda_{ex}$ = 405 nm and $\lambda_{ex}$ = 340 nm). The crystal structure of the NCs was determined by performing X-ray diffraction (XRD) analysis using a Bruker D8 Advance diffractometer and a Rigaku Smartlab equipped with Cu-$k_\alpha$ (1.54 Å) radiation. Microscopic imaging was conducted using a JEOL JEM-2100 Plus transmission electron microscope, operating at an accelerating voltage of 200 kV, and employing the bright field imaging technique. To prepare the samples, a carbon-coated Cu grid was utilized, onto which a droplet of purified NCs dissolved in hexane was placed. Subsequently, the solvent was allowed to evaporate, leaving the NCs ready for imaging. At room temperature, the electron paramagnetic resonance (EPR) spectra for NCs were obtained using a JEOL Resonance JES-X320 ESR Spectrometer, operating in the X-band (9.34 GHz). The quantitative determination of the dopant concentration ($Mn^{2+}$) was achieved by analysing the elemental composition of the samples after acid digestion, utilizing an inductively coupled optical emission spectrometer (PerkinElmer ICP-OES Avio 220 Max). We conducted MCD measurements using a customized setup that involved combining a CD spectrometer with a magnet. The light source, which could be tuned to different wavelengths, utilized a 450 W water-cooled Xe lamp and a double prism polarizing monochromator (JASCO J-1500-450). For the experiment, the sample was drop cast onto a spectrosil B quartz slide and positioned between two poles of the spectromag PT, which incorporated a 7T split pair solenoid magnet from Oxford Instruments. The setup followed the Faraday geometry and included a cryostat for temperature-dependent measurements. The polarized light passed through the sample, and the transmitted light was detected by a photomultiplier.

**Results and discussions**

A modified version of the synthesis method reported by Protesescu et al.[26] was used to prepare Mn-doped $CsPbCl_3$ (MnP1) NCs. Various amounts of dopant concentration were obtained by altering the Pb/Mn stoichiometry during the reaction, as determined through ICP-OES analysis. Experimental section provides a detailed description of the synthesis procedure and includes the sample code. The as-prepared NCs of $CsPbCl_3$ (P1) display absorption and PL characteristics similar to previous reports,[26-28] exhibiting narrow band-edge emission at 3.06 eV (Figure 1a). In the doped NCs, the bandgap emission shows a slight blue shift compared to the undoped counterparts, and there is an additional broad PL feature at around 2.1 eV, which is attributed to $Mn^{2+}$ *d-d* emission (blue and red curve in Figure 1a). Importantly, the gated PL collected after a few microseconds of delay confirms the presence of long-lived Mn emission. The gated PL excitation (PLE) spectrum, obtained by monitoring PL at 2.1 eV, closely matches the absorption spectrum, indicating the sensitization of Mn emission from the P1 host (Figure 1b). Additionally, the inset in Figure 1b demonstrates the millisecond-range lifetime of Mn emission, further confirming the successful integration of Mn into the lattice. Figure 1c shows powder XRD pattern indicating the formation of NCs in cubic phase with high crystallinity and purity. Peak broadening in comparison to bulk $CsPbCl_3$ suggests the formation of NCs. Further EPR spectroscopy was chosen to characterize $Mn^{2+}$ speciation. Figure 1d displays the X-band EPR spectra of colloidal P1 doped with 2.1% and 6.9% Mn, measured at room temperature. The experimental spectrum exhibits a distinct six-line hyperfine splitting pattern of $Mn^{2+}$ (I=5/2), indicating an approximate spacing of ~86 Gauss. This implies that the environment surrounding Mn is octahedral and it occupies the Pb position in the perovskite structure. Moreover, formation of NCs was verified through TEM analysis. Figure 1e and f display a cubic shape, with average sizes of 11.4 ± 1.4 nm for P1 and 9.1 ± 0.9 nm for MnP1 (Mn 2.1%), respectively. All these studies validate the inclusion of Mn into the perovskite lattice, and the transfer of electron from host to a dopant result in a stoke-shifted emission at 2.1 eV.

As depicted in Figure 1, this study provides compelling evidence for nanocrystal doping. The observation of Mn emission suggests the existence of the anticipated exciton-Mn energy transfer resulting from the interaction of *sp-d* exchange. However, they do not serve as a conclusive illustration of *sp-d* exchange interactions within these QDs. MCD spectroscopy was employed to measure the *sp-d* exchange interactions in these QDs, specifically by studying the excitonic Zeeman splitting. First the thin films of these NCs were prepared by drop-cast onto a quartz substrate for magneto-optical measurement using MCD spectrometer. It is widely recognized in QDs, with their high surface-to-volume ratios, can display stronger *sp-d* exchange interactions compared to their bulk counterparts.[29-30] Figure 2 (a-b) depicts the 1.5 K absorption and 5T MCD spectra of P1 and MnP1(2.1%) NCs. Both of these samples display multiple well-resolved excitonic transitions in their absorption and MCD, providing a distinct opportunity to thoroughly investigate the photo physics of the excited states. The first three excitonic transitions are denoted as (following the assignment as described in detail by Mandal et al.[31]) X$_1$ ($1s_h - 1s_{so\ J=1/2}$), X$_2$ ($1s_h - 1s_{e\ J=3/2}$), and X$_3$ ($1p_h - 1p_e$), (indicated by a dotted line), respectively. The precise determination of the energy position for each transition is accomplished by identifying the peaks of the first derivative of the MCD spectra (represented by the blue curve in 2 a-b). The effect of a magnetic field was investigated by collecting variable field MCD spectra ranging from 0 to 5 T for P1 and MnP1 (2.1%) at a temperature of 1.5 K, as shown in Figure 2c, d. It is apparent from the spectra that there are distinct variations in the MCD signal between doped QDs and P1 for all the transitions. Additionally, the Zeeman splitting energy ($\Delta E_z$) was calculated using equation 1 in order to quantify all of these changes.

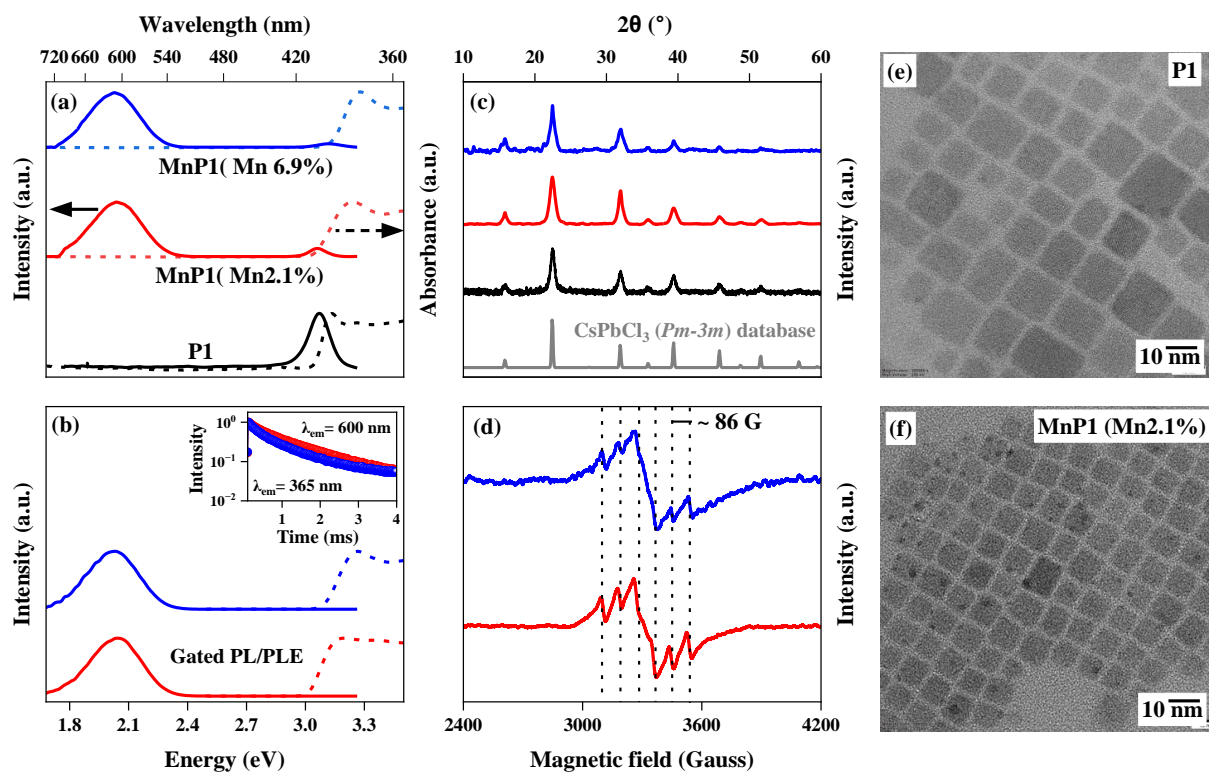

**Figure 1** Basic characterization of undoped CsPbCl$_3$ (P1) and Mn-doped CsPbCl$_3$ (MnP1). (a) Optical absorption (dotted lines) and emission spectra (solid lines) of P1 and MnP1, (b) Gated PL and PL excitation spectra of the Mn-doped sample. In the inset lifetime of the Mn-emission ($\lambda_{em}$ = 600 nm) obtained with $\lambda_{ex}$ = 365 nm, (c) X-ray diffraction pattern along with the bulk data from ICSD (d) Electron paramagnetic resonance (EPR) signal of the Mn-doped samples. (e, f) TEM images of undoped and doped NCs.

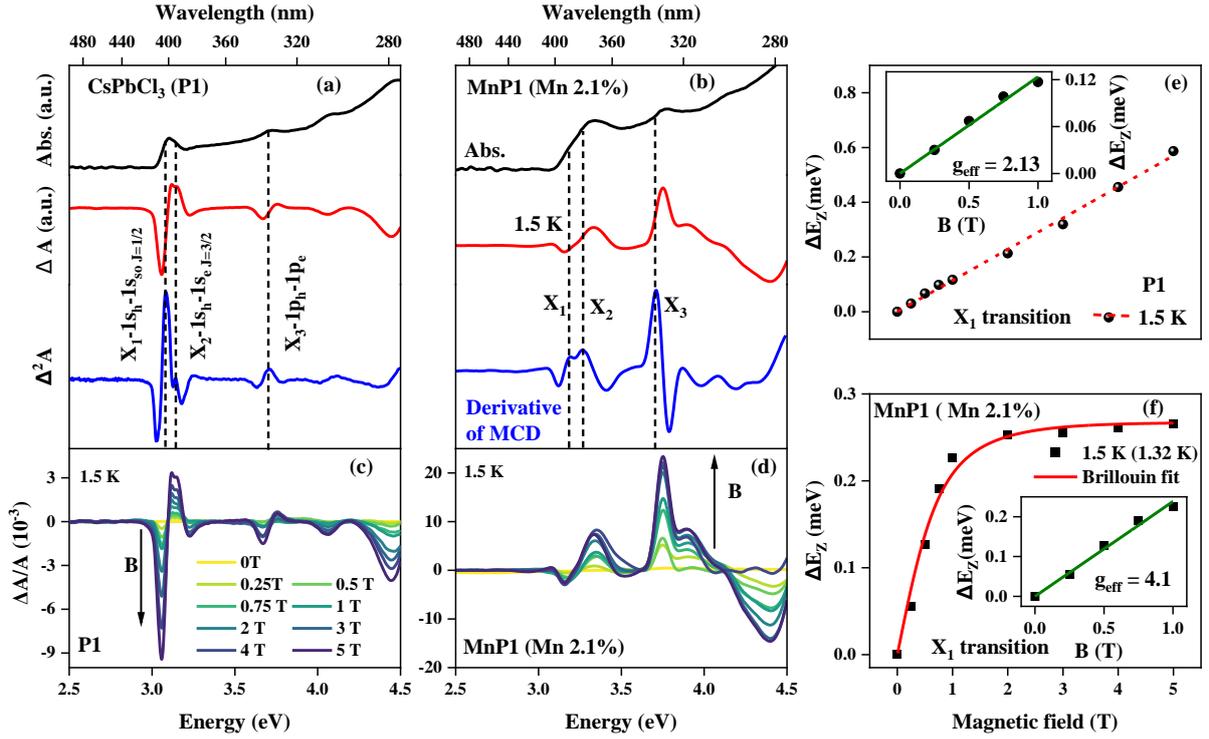

**Figure 2** Magneto-optical study of undoped $CsPbCl_3$ (P1) and Mn-doped $CsPbCl_3$ with 2.1 % of Mn (Mn 2.1%) by magnetic circular dichroism (MCD) spectroscopy. (a) Absorption spectra of P1 obtained from the MCD spectrometer at 1.5 K temperature and 0 T magnetic field (black curve) and MCD spectra at 1.5 K with 5T magnetic field (red curve). The Maxima of the 1st derivative of MCD spectra (blue curve) are taken to find out the energy of the excitonic transitions and marked as $X_1$, $X_2$, and $X_3$ for the first three transitions, respectively. (b) Absorption, MCD, and derivation of MCD spectra for 2.1 % Mn-doped P1 sample. MCD spectra at 1.5 K with the variance of the magnetic field from 0.25 T to 5T for (c) P1 and (d) 2.1 % Mn-doped P1. (e) Zeeman splitting energy ($\Delta E_Z$) as a function of magnetic field for $X_1$ transition at 1.5 K for P1. Red dashed line is just a guide to the eye. (f) Zeeman splitting energy ($\Delta E_Z$) as a function of magnetic field for $X_1$ transition at 1.5 K for MnP1 (Mn 2.1%). Red line is the Brillouin fit to the Zeeman energy. Inset of (e & f) showing linear fit to the Zeeman energy up to 1T magnetic field to determine the *g*-effective value.

**Equation 1**[32] $$\Delta E_z\ (MCD) = -\left(\frac{\sqrt{2e}}{2}\right)\sigma\frac{\Delta A}{A_0}$$

where $\Delta A = \frac{\theta\ (mdeg)}{32980}$, $\Delta A$ is the maximum amplitude lowest-energy leading-edge of the MCD feature, $A_0$ is the absorption at the maxima of the MCD peak, $\sigma$ is the Gaussian width of the peak.

The true defining attribute of a DMS QDs MCD intensities is the magnetic saturation behaviour upon varying magnetic field. The $\Delta E_z$ of a doped QDs can be described by using equation 2. The first term in the equation represents the relatively small intrinsic excitonic Zeeman splitting ($\Delta E_{int}$). Here, $g_{int}$ denotes the intrinsic g value of the exciton, $\mu_B$ represents the Bohr magneton, and B represents the external magnetic field. The second term encompasses the contributions arising from the *sp-d* exchange (equation 3) which have contribution from the effective Mn$^{2+}$ concentration, exciton-dopant overlap (γ), dopant *sp-d* exchange contribution parameter $N_0(\alpha - \beta)$, and $\langle S_Z \rangle$ the spin expectation value of Mn$^{2+}$, which follows Brillouin behaviour for a spin-only S = 5/2 (equation 4) ground state. The intensity of the X$_1$ transitions in the MCD spectrum (Figure 2c) of P1 exhibits a progressively negative low-energy leading-edge intensity as the magnetic field increases. This indicates a positive Zeeman splitting of P1, as depicted in Figure 2e and its follows nearly linear dependency with the magnetic field as shown by the red dotted line. Conversely, the same peak intensity rises at lower magnetic fields and reaches a saturation point at higher magnetic fields, as illustrated in Figure 2f for MnP1 (Mn 2.1%). These Zeeman energy data fitted to the equation 4 with $g_{Mn} = 2.0042$, $k_B$ is the Boltzmann constant, $\gamma = 1$, and T= experimental temperature. The data exhibits Brillouin-like behaviour in paramagnetic $Mn^{2+,}$ with a linear increase in the low-field (Curie) regime followed by saturation at higher magnetic fields and also the fitted temperature (1.32K) matching with the experimental value. The typical Brillouin function fitting (equation 4) of MnP1 (Mn 2.1%) excitonic MCD intensities provide clear evidence of exciton-$Mn^{2+}$ exchange coupling in these doped QDs. Moreover, calculations of the effective exciton $g$-factor at magnetic fields below one tesla indicate a significant rise in $g_{eff}$ from 2.13 (for P1) to 4.1 (for MnP1(2.1%)) for the $X_1$ transitions, as demonstrated in the inset of Figure 2 (e-f).

**Equation 2** $$\Delta E_z = \Delta E_{int} + \Delta E_{sp-d}$$

**Equation 3** [33] $$\Delta E_z = g_{int}\mu_B B + x_{eff}\gamma N_0(\alpha - \beta)\langle S_Z \rangle$$

**Equation 4** [33]

$$\Delta E_{sp-d} = x_{eff}\gamma N_0(\alpha - \beta)\left[\frac{2S+1}{2}coth\left(\frac{2S+1}{2}\frac{g_{Mn}\mu_B B}{K_B T}\right) - \frac{1}{2}coth\left(\frac{g_{Mn}\mu_B B}{K_B T}\right)\right]$$

Upon Mn doping, these results indicate that the $\Delta E_z$ and corresponding $g_{eff}$ value for the band edge transition in perovskite QDs do not show a significant change, in contrast to an earlier report on II-VI DMS QDs. In the case of $Mn^{2+}$ and $Co^{2+}$ doped CdSe QDs, the earlier report suggests that $\Delta E_z$ is approximately 9-10 meV, and $g_{eff}$ is around 50.[20] The fundamental difference arises from the fact that perovskite has an inverted band structure, where the split-off state is present in the conduction band and contributes to the band edge transition. However, in II-VI semiconductors, the split-off state is not directly involved in the band edge transition as it resides in the valence band. The schematic band diagram of perovskite is shown in Figure 3a. Though for MnP1 (2.1%), the band edge transition shows Brillouin-like behaviour at 1.5 K, at higher temperatures, its behaviour becomes random, as shown in Figure 3b. Barrows et

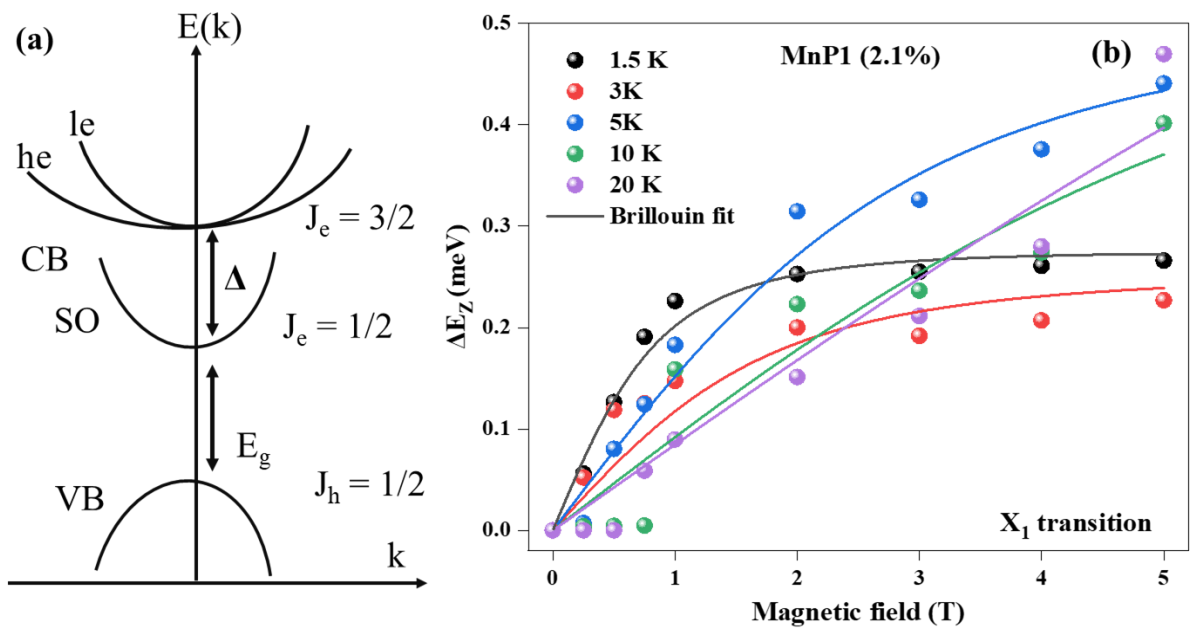

**Figure 3** (a) Schematic band diagram of perovskite. (b) Temperature dependent Zeeman splitting energy ($\Delta E_Z$) as a function of magnetic field for $X_1$ transition MnP1(2.1%). Solid line is the Brillouin fit to the Zeeman energy.

al. demonstrate that the extent of the *sp-d* exchange interaction can be manipulated by controlling the level of Mn impurities in CdSe QDs.[33] Previously, in the case of Mn-doped CdSe, the Zeeman splitting of the upper excited states was observed as a result of valence band mixing.[34] A recent study on electronic structure, investigating the back transfer mechanism in Mn-doped perovskite, has unveiled that the coupling of the Mn-state occurs with a higher excited state rather than the first excited state.[17] Therefore, in the case of perovskite as well, there is always a finite probability of *sp-d* exchange interaction occurring in the upper excited state and synthetic engineering also allows for the manipulation of Mn impurity level. This control enables the regulation of the spatial overlap between the dopants and the confined charge carriers within the semiconductor. Compositional variation, variation in dopant concentration, and the contribution of higher excited state Zeeman effect can provide a comprehensive understanding of the *sp-d* exchange interaction in these perovskite QDs.

Therefore, we synthesized Mn-doped P1, which contained a higher Mn concentration of approximately 6.9%. The particle size was determined to be 8.1 ± 0.7 nm through TEM analysis. Then, we investigated the characteristics of its higher excitonic transitions, namely $X_2$ and $X_3$, in comparison to undoped and less Mn-doped QDs. Figure 4a-c illustrate the Zeeman splitting energy for the $X_2$ transition of P1, MnP1 (Mn 2.1%), and MnP1 (Mn 6.9%), respectively as a function of temperature and magnetic field. For undoped QDs, $\Delta E_Z$ is nearly independent of temperature and corresponding $\Delta E_Z$ is very small (few μeV) as shown in Figure 4a. The $\Delta E_Z$ value increases significantly by a few meV upon Mn doping and exhibits temperature-dependent behaviour (Figure 4 b-c). We have observed that an increased concentration of Mn results in a greater Zeeman splitting proportional to the concentration.

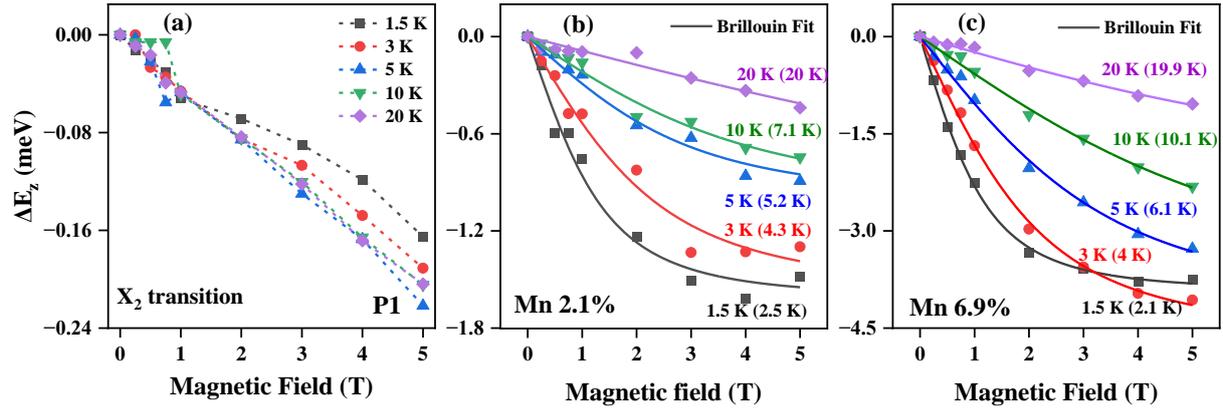

**Figure 4** Comparison of Zeeman splitting energy ($\Delta E_Z$) for the $X_2$ transition in P1 and various % of Mn-doped MnP1 NCs. (a) $\Delta E_Z$ as a function of temperature (1.5 K to 20K) and magnetic field for P1 NCs and dotted lines are just guided to the eye. (b) $\Delta E_Z$ as a function of temperature (1.5 K to 20K) and magnetic field for MnP1 (Mn 2.1%) NCs and solid lines are Brillouin fit to Zeeman energy. Fitted temperatures are written in the bracket. (c) $\Delta E_Z$ as a function of temperature (1.5 K to 20K) and magnetic field for MnP1 (Mn 6.9%) NCs and solid lines are Brillouin fit to Zeeman energy. Fitted temperatures are written in the bracket.

Equation 4 is used to fit $\Delta E_Z$ vs $B$ plot in 4 (b-c) revealing a Brillouin-like magnetic saturation behaviour resulting from the strong *sp-d* exchange interaction of the paramagnetic $Mn^{2+}$ ions with the host charge carriers also exhibiting temperature dependence. Fitted temperature (shown in bracket) also closely matched with the experimental value. This implies a strong coupling between Mn and the second excited state of the host, resulting in the emergence of a significant Zeeman splitting. Additionally, the temperature increases lead to a reduction in $\Delta E_Z$, suggesting a decreased occupancy of the Zeeman sublevel at higher temperatures. Additionally, we have determined the $g_{eff}$ value by linear fit to the equation $\Delta E_z = g_{eff}\mu_B B$ for $\Delta E_Z$ vs $B$ plot at lower magnetic field strengths as shown in Figure 5. Significantly, there was an observed notable increase in the $g_{eff}$ value, which transitioned from (-) 0.8 (for P1) to (-)14.2 (for Mn 2.1%) and eventually reached (-) 41.4 (for Mn 6.9%). Thus, carrier spin can be manipulated by varying the amount of Mn doping in perovskite QDs.

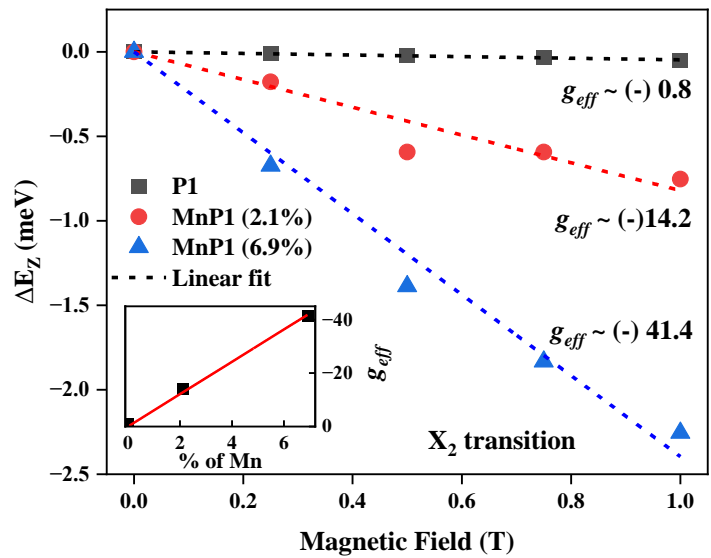

**Figure 5** Linear fit to the equation $\Delta E_z = g_{eff}\mu_B B$ for $\Delta E_Z$ vs B curve at 1.5 K up to 1T magnetic field to determine the effective g-factor for the $X_2$ transition in P1 and Mn doped QDs.

Based on the aforementioned observation, we conducted additional investigation on the third excitonic transition (referred to as $X_3$) to gain a deeper understanding. Figure 6a illustrates the representative MCD spectra of MnP1 (6.9%) at different temperatures while keeping the magnetic field constant. The spectra clearly indicate a decrease in MCD intensity as the temperature rises. However, the Zeeman splitting energy of the $X_3$ transition in these doped NCs is compared with that of P1 and illustrated in Figure 6 (b-c). For the transition of $X_3$, the sign of the MCD signal changes upon Mn-doping, indicating the differing polarizability of the intrinsic and *sp-d* contributions, as illustrated in Figure 6b. The intrinsic contribution is negligible and remains temperature-independent, as indicated by the dotted line for P1. The data for the 2.1% and 6.9% Mn-doped samples, along with their fit to equation 4, suggest a strong Brillouin-like behaviour of paramagnetic $Mn^{2+}$ ions, accompanied by magnetic saturation at higher magnetic fields as shown in Figure 6b and 6c respectively. In this case, the fitted temperature also matches well with the experimental data. In the case of these doped NCs, it has been observed that the Zeeman splitting energy is 1000 times higher when compared to undoped NCs, which is an unprecedented increase. Interestingly, there is no existing report on perovskite QDs that mentions such a significant boost in Zeeman splitting upon Mn-doping. We performed a fitting of the data to the equation $\Delta E_z = g_{eff}\mu_B B$ for the $\Delta E_Z$ vs B curve at 1.5 K up to 1T. We were able to extract the *g*-factor for the $X_3$ transition and the results are shown in Figure 7. The graph clearly indicates a change in sign and relative magnitude of the effective exciton *g*-factor upon Mn-doping for this transition. For the undoped case, the slope is positive, and the $g_{eff}$ value is 0.11. However, for the doped sample, the slope is negative, and the $g_{eff}$ value varies depending on the dopant concentration, reaching values as high as (-) 313.7 (Figure 7b) Hence, in these perovskite NCs, the *sp-d* exchange not only contributes to the bandgap state, akin to II-VI QDs, but also strongly interacts with the higher excited state. Notably, the nature of this higher-order transition differs from the band edge state, as evidenced by the reversal of the $\Delta E_z$ sign.

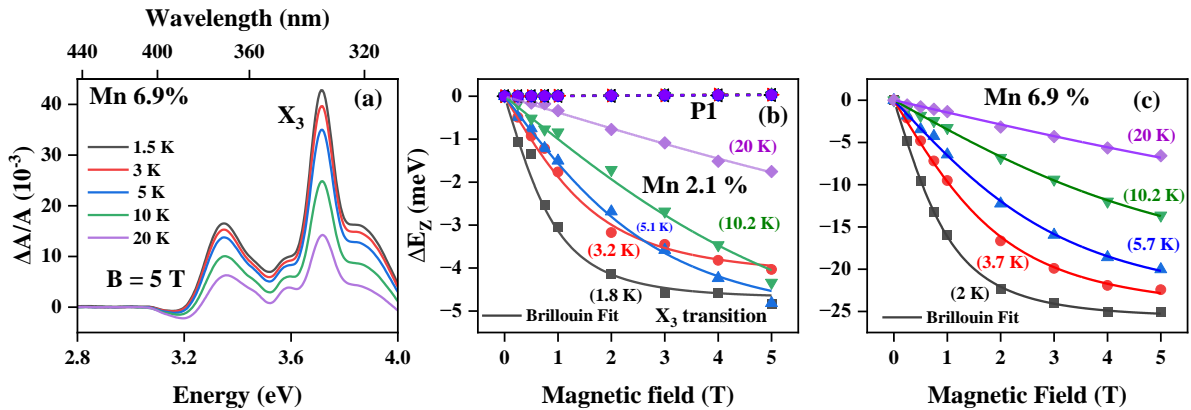

**Figure 6** Comparison of Zeeman splitting energy ($\Delta E_Z$) for the $X_3$ transition in P1 and various % of Mn-doped MnP1 NCs. (a) MCD spectra as a function of temperature at a fixed magnetic field of 5 T for MnP1 (Mn6.9%) NCs. (b) $\Delta E_Z$ as a function of temperature ((1.5 K to 20K) and magnetic field for P1 and MnP1 (Mn 2.1%) NCs. Dotted lines are guided to the eye showing the $\Delta E_Z$ for P1 NCs. For MnP1 (Mn 2.1%), solid lines are the Brillouin fit of the $\Delta E_Z$ vs. B curve. Fitted temperatures are written in the bracket (c) $\Delta E_Z$ as a function of temperature (1.5 K to 20K) and magnetic field for MnP1 (Mn 6.9%) NCs and solid lines are Brillouin fit to Zeeman energy. Fitted temperatures are written in the bracket.

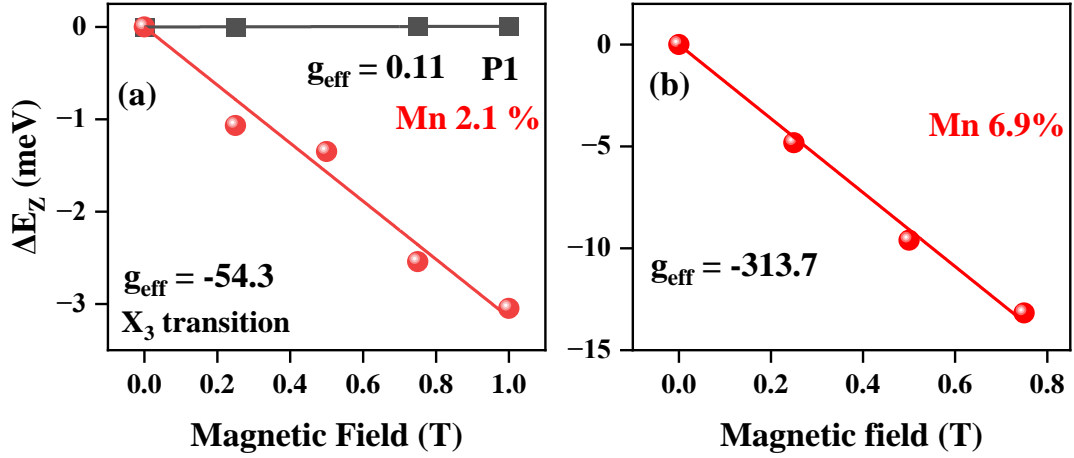

**Figure 7** Zeeman energy versus magnetic field in the lower magnetic field (to find out the exciton *g*-factor, i.e., $g_{eff}$ ). Comparison of exciton *g*-factor for the X$_3$ transition. (a) P1 and MnP1 (with Mn 2.1%). Solid liners are linear fit, and $g_{eff}$ is calculated from the slope. (b) MnP1 (Mn 6.9%).

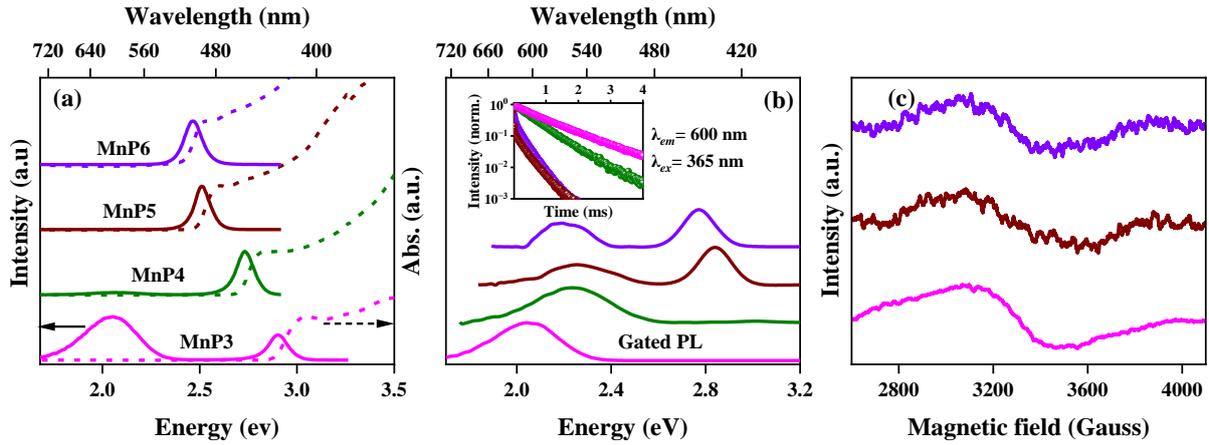

**Figure 8** Basic characterization of Mn-doped NCs with varying halide composition (MnP3-MnP6). (a) Optical absorption (dotted lines) and emission spectra (solid lines), (b) Gated PL and Mn-lifetime ($\lambda_{em} = 600$ nm, $\lambda_{ex} = 365$ nm) in the inset. (c) EPR signal from the corresponding samples.

Once we have gained a comprehensive understanding of the Zeeman interaction in chlorine-based perovskite QDs, our objective is to investigate the impact of Mn doping with varying halide composition. To accomplish this, we have successfully synthesized Mn-doped

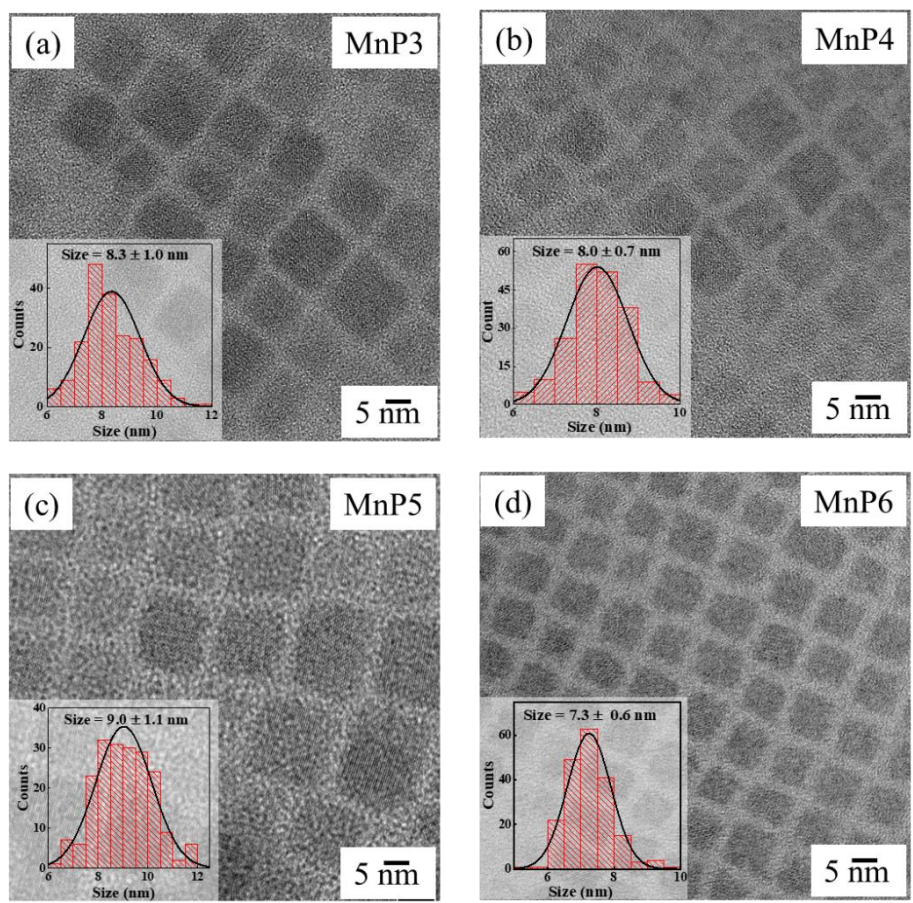

**Figure 9** TEM images along with the histogram of size analysis in the inset. (a) MnP3 (8.3 ± 1.0 nm), (b) MnP4 (8.0 ± 0.7 nm), (c) MnP5 (9.0 ± 1.1 nm), and (d) MnP6 (7.3 ± 0.6 nm).

QDs comprising a mixture of chlorine and bromine, as well as pure bromine based QDs, utilizing the methodology outlined in the experimental section. Figure 8a depicts the absorption and steady-state PL spectra of MnP3-MnP6 at room temperature, with the compositions provided in Table T1. Similar to previous reports, the incorporation of Br into the NCs causes a shift in the exciton PL towards the lower energy range. In addition, a broad emission at 590 nm (2.1 eV), arising from the spin-forbidden $^4T_1$ to $^6A_1$ transition of the Mn-d levels is observed in samples with larger bandgap ($E_g$). However, in cases where $E_g$ of perovskite closely matches the excitation energy of Mn (~500 nm), no Mn emission is observed,[35] like the scenario with MnP5 and MnP6 here. Further, the samples were characterized by measuring gated emission, where excitonic emission was obtained after a delay of 200 $\mu s$, and Figure 8b showcases typical gated emission spectra at room temperature for various halide concentrations.

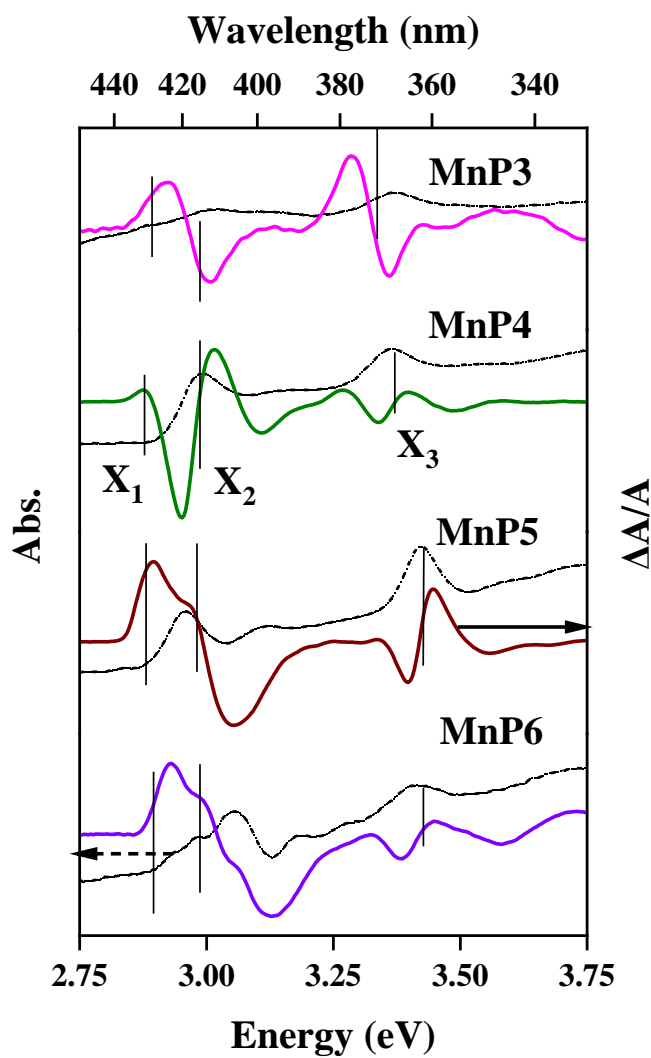

**Figure 10** Optical absorption in thin film at 1.5 K and MCD spectra at 1.5 K with a magnetic field of 5T for MnP3-MnP6 NCs. Absorption spectra are shown with the dotted lines and MCD with the solid lines. The first three transitions ($X_1$, $X_2$, and $X_3$) are marked with a line. Here we have matched the bandgap energy with MnP3 by adding a constant term ($\Delta E$) for the MnP4-MnP6 NCs. $\Delta E$ = 0.081, 0.317, and 0.412 eV for MnP4, MnP5, and MnP6 respectively.

From the figure, it is evident that all samples show Mn emission, indicating that Mn possesses a long-lived lifetime in the order of milliseconds, as depicted in the inset of 8b. However, MnP5 and MnP6 exhibit band edge emission similar to those reported earlier,[36] which can be attributed to electron back transfer originating from Mn. Furthermore, broad EPR signal shown in Figure 8c obtained from these samples because of the merging of individual signals may be attributed to the presence of Mn-Mn antiferromagnetic interactions. These studies collectively validate the successful integration of Mn into the mixed halide perovskite QDs. Quantitative analysis of the elemental composition revealed Mn concentrations of approximately 2.7%, 2.2%, 1.4%, and 1.5% in MnP3, MnP4, MnP5, and MnP6, respectively. Figure 9 depicts high-resolution TEM images of MnP3-MnP6 NCs. The nanocrystals exhibit a cubic shape and possess an average size of 7 to 9 nm.

After characterizing the mixed halide-based samples, we proceeded to examine their magneto-optical properties through MCD spectroscopy. In Figure 10, the absorption spectra at 1.5 K are represented by the black dotted line, and the corresponding 5T MCD spectra for MnP3-MnP6 are displayed. The first three transitions were identified by locating the peaks of the first derivative of the MCD spectra (not shown). To plot the data, we include a constant ΔE in the $E_g$ value of MnP3, which corresponds to the variation in $E_g$ resulting from changes in compositions, to depict the absorption and MCD spectra of MnP4-MnP6 NCs. Our observations indicate that although the $X_2$ transition energy remains unaffected by changes in composition but the $X_3$ transition shifts towards higher energy as the Br content increases.

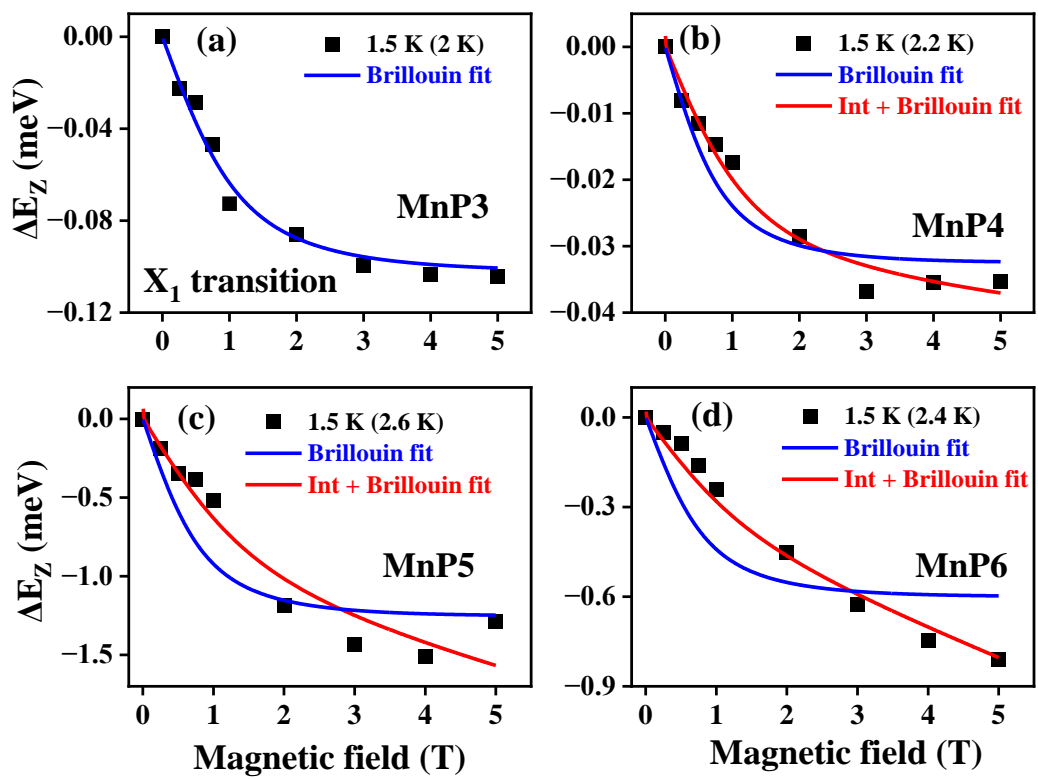

**Figure 11** The plot of $\Delta E_Z$ as a function of the magnetic field, along with the corresponding fit for the $X_1$ transition in MnP3-MnP6. The blue solid line represents the Brillouin fit, while the red solid line corresponds to the intrinsic plus Brillouin fit. The fitting temperatures are indicated in brackets.

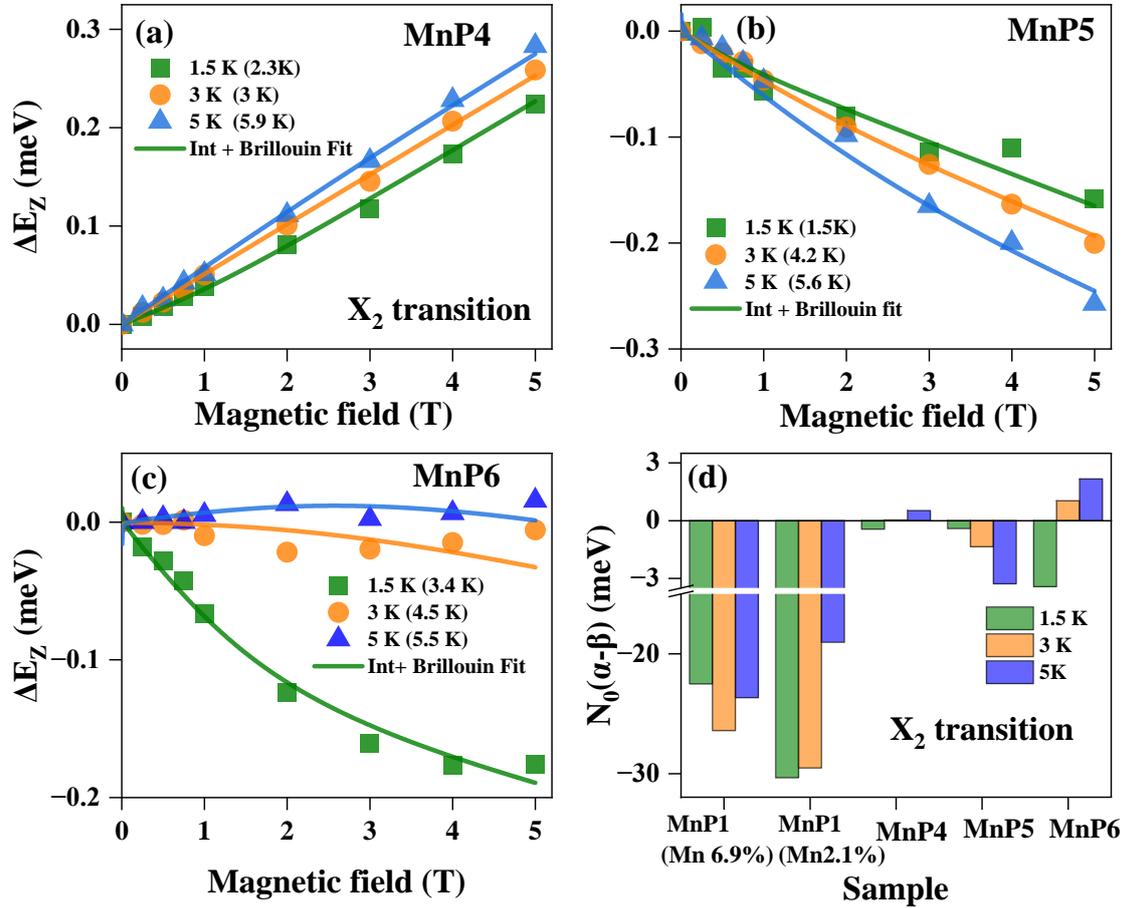

**Figure 12** Probing the $X_2$ transitions. (a-c) The plot of $\Delta E_Z$ as a function of the magnetic field and temperature along with the corresponding fit (solid lines) using equation 3 considering both intrinsic and *sp-d* contribution. Fitted temperature are shown in brackets. (d) Comparison of $N_0(\alpha - \beta)$, obtained from the fitting, for these Mn-doped QDs.

As the $E_g$ decreases, the Mn emission intensity decreases as well, and in pure MnP6, there is no emission (shown in Figure 8a). However, at room temperature, back transfer is apparent from the gated PL measurement (Figure 8b), as we do observe PL from the band edge for all the samples. As we have already observed Mn couples with the band edge and higher excited state through *sp-d* exchange interaction, resulting in a significant Zeeman splitting in the case of MnP1. This raises an open question regarding the coupling mechanism between Mn and the host in these materials. Therefore, a thorough investigation was conducted on each transition by collecting MCD data at different temperature and magnetic field values. Figure 11 illustrates the variation of Zeeman energy with the magnetic field for the $X_1$ transition at a temperature of 1.5 K. In the case of MnP3, although the Zeeman energy is relatively low, the data fits well with the Brillouin function described in equation 4. This suggests a weak *sp-d* exchange interaction. However, when we consider lower band gap materials, the data cannot be adequately fitted using only the Brillouin function but can be fitted considering intrinsic contribution using equation 3 as shown in Figure 11 (b-d) for MnP4 to MnP6. This discrepancy arises because the intrinsic contribution and *sp-d* contribution become comparable in these cases. As we observed in the case of MnP1, Mn d-electrons mostly interact with the higher-order excitonic states, resulting in a giant Zeeman splitting. Therefore, we conducted investigations on the $X_2$ and $X_3$ transitions for these samples. Figure 12 (a-c) and 13 (a-c) illustrates the change in $\Delta E_Z$ with respect to temperature and magnetic field in MnP4 - MnP6 for the $X_2$ and $X_3$ transition respectively. The data indicates that $\Delta E_Z$ is not as significant as observed in the case of MnP1.

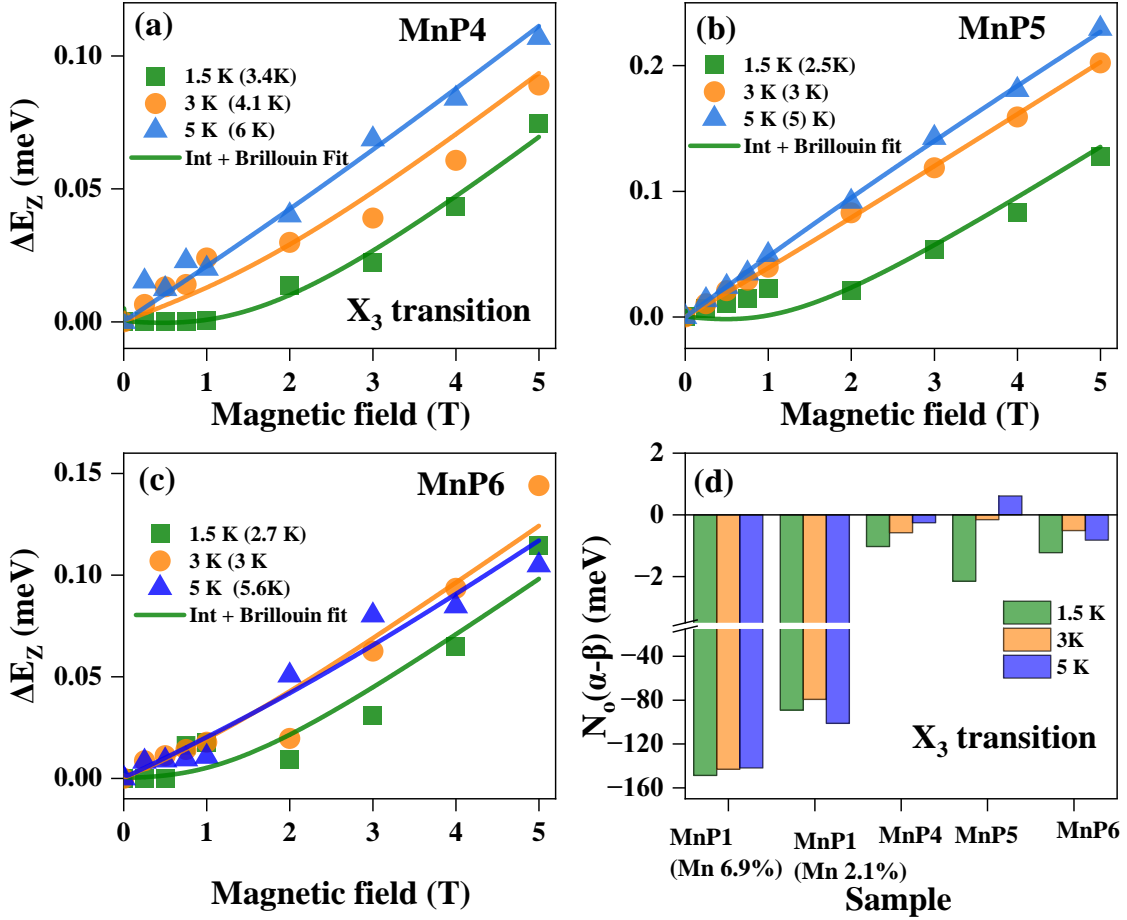

**Figure 13** Probing the $X_3$ transitions. (a-c) The plot of $\Delta E_Z$ as a function of the magnetic field and temperature along with the corresponding fit (solid lines) using equation 7.3 considering both intrinsic and *sp-d* contribution. Fitted temperature are shown in brackets. (d) Comparison of $N_0(\alpha - \beta)$, obtained from the fitting, for these Mn-doped QDs.

**Table T2** Comparison of $g_{eff}$ in Undoped and Doped Perovskite vs II-VI QDs

| Sample | Method | $|g_{eff}|$ (Temp.) | $N_0(\alpha - \beta)$ (eV) | Ref. |
|---|---|---|---|---|
| $Mn^{2+}$: CdSe | MCD | 52.5 (6K) | $+1.37^{37}$ (bulk) | Archer et al.[20] |
| $Mn^{2+}$: CdSe | MCD, MCPL | 300 (6K) | $+1.37^{37}$ (bulk) | Beaulac et al.[38] |
| $Mn^{2+}$: ZnSe/CdSe | MCD | 200 (1.6K) | - | Bussian et al.[39] |
| $Mn^{2+}$: CdSe | MCD | 907 (1.8K) | - | Vlaskin et al.[40] |
| $CsPbCl_3$ | MCD | 2.1 (1.5 K) | - | This work |
| Mn: $CsPbCl_3$ (Mn 2.1%) | MCD | 4.1 (1.5 K) – ($X_1$)<br>14.2 (1.5 K)-$X_2$<br>54.3 (1.5 K) -$X_3$ | +0.005<br>-0.030<br>-0.089 | This work |
| Mn: $CsPbCl_3$ (Mn 6.9.1%) | MCD | 54.3 (1.5 K)-$X_2$<br>314 (1.5 K) -$X_3$ | -0.022<br>-0.148 | This work |

Nonetheless, a minimal *sp-d* interaction results in a small splitting that is comparable to the intrinsic splitting. The data fit well using equation 3, and the temperature used for fitting matches the experimental value. The weak splitting observed is a result of the weak interaction between the charge carriers and localised Mn-spin, which can be determined by the value of $N_0(\alpha - \beta)$. We acquired this value through fitting and represented it in Figure 12d and 13d for the $X_2$ and $X_3$ transition respectively. In the case of MnP1, $|N_0(\alpha - \beta)|$ is relatively high, ranging from 20-30 meV for $X_2$ and 90-120 meV for $X_3$ transitions. However, for MnP4-MnP6, it is only in the range of 1-2 meV. As a result, when introducing Mn into materials that possess a low band gap, we observe a minimal Zeeman splitting. However, the observed sign changes in $\Delta E_Z$ with temperature in the case of MnP6 for the $X_2$ transitions can be explained by the variation in the exchange interaction parameter value, $N_0(\alpha - \beta)$, which changes from (–)3.4 meV at 1.5 K to 2.1 meV at 5 K. Therefore, at 1.5 K, both the intrinsic and *sp-d* contributions have the same sign. At 3 K, they oppose each other and almost cancel out. Finally, at 5 K, $\Delta E_{Z\ (sp-d)}$ slightly dominates, resulting in a change in the sign of $\Delta E_Z$. For $X_3$ transition $\Delta E_Z$ is dominated by the intrinsic contribution for the MnP4-MnP6 QDs. In Table T2, we have summarized our results and compared them with those of Mn-doped II-VI semiconductor QDs. This report represents the first instance in which we have experimentally calculated the $g_{eff}$ and $N_0(\alpha - \beta)$ value in $Mn^{2+}$-doped perovskite QDs. In perovskite QDs, the $g_{eff}$ value we achieved can exceed 300 but still lower than diffusional Mn -doped CdSe ($|g_{eff}| = 907$).[40] The lower value of $N_0(\alpha - \beta)$ may be attributed to the distinct band structure of perovskite QDs compared to II-VI QDs.

In order to elucidate the aforementioned observations, a basic schematic 1 is presented to illustrate the Zeeman splitting mechanism. Upon application of magnetic field, the VB and CB state no longer remain degenerate. They split according to their intrinsic *g*-value. Now application of LCP and RCP light absorb differentially and give rise to Zeeman splitting which can be accounted using $\Delta E_Z = g\mu_B B$ in case of undoped QDs. Thus $\Delta E_Z$ of diamagnetic undoped perovskite QDs is linearly proportional to the applied magnetic field strength as shown in Figure 2e. The introduction of doping results in the exchange coupling between the exciton and magnetic dopants, which leads to a new contribution to the excitonic Zeeman splitting energy. This contribution relies on the strengths of two types of $Mn^{2+}$-carrier exchange interactions: $N_0\alpha$ (*s-d*, or potential exchange) and $N_0\beta$ (*p-d*, or kinetic exchange), as well as the effective concentration of $Mn^{2+}$ ions ($x$). Under the influence of an external magnetic field, the spins of $Mn^{2+}$ become aligned, and their exchange energies combine additively. The observation reveals a contribution to the alteration in $\Delta E_Z$ that is specific to the dopant. Hence, the value of $\Delta E_Z$ is contingent upon the anticipated spin value $\langle S_Z \rangle$ exhibited by the $Mn^{2+}$ dopant in alignment with the direction of the magnetic field that is applied. Thus, the splitting in the CB and VB is enhanced by two orders of magnitude compared to the undoped perovskite QDs. However, altering the composition of the pristine

QDs weakens the coupling strength from MnP1 to MnP6, resulting in smaller splitting and consequently a smaller value of $\Delta E_Z$.

**Schematic 1** Zeeman splitting of valence band (VB) and conduction (CB) in presence of magnetic field in undoped and Mn-doped perovskite QDs.

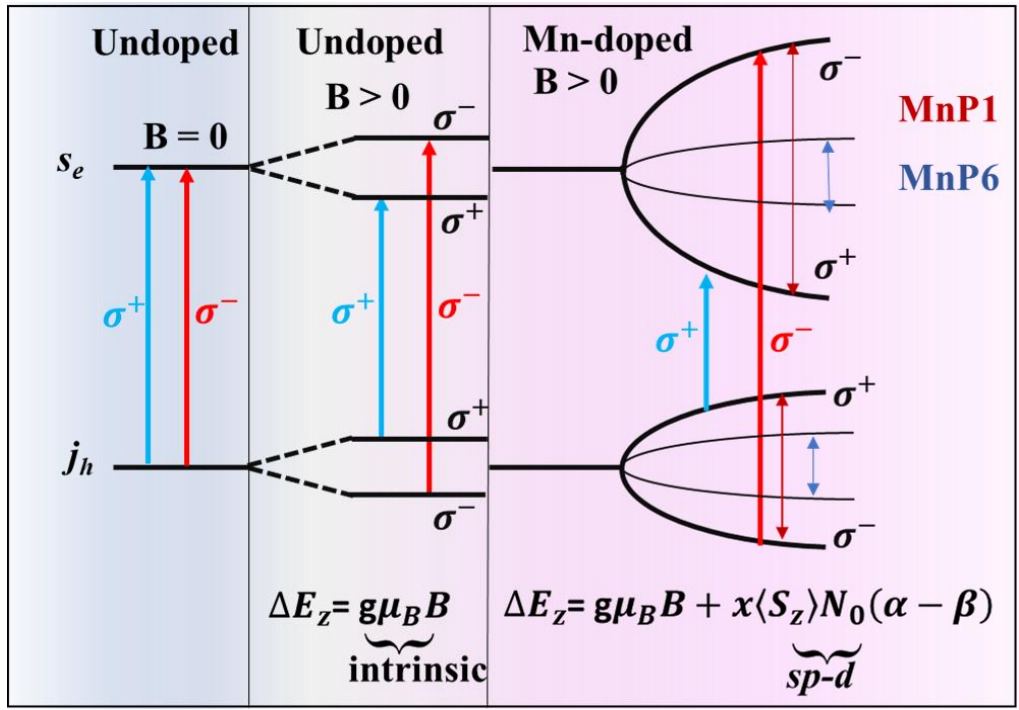

**Conclusions**

In summary, the effective utilisation of MCD spectroscopy has demonstrated the presence of strong *sp-d* exchange interactions in colloidal $Mn^{2+}$-doped $CsPbCl_3$ NCs, whereas feeble interactions have been detected in $CsPb(Cl/Br)_3$ and $CsPbBr_3$ NCs. In this study also revealed that Mn -coupled strongly with the higher excited state which facilitate the electron back transfer from Mn to host reported elsewhere.[17, 36] The utilisation of MCD spectroscopy facilitated the assessment of the excitonic Zeeman splitting. According to the results, the value of $g_{eff}$ was estimated to be ~ 314 at a temperature of 1.5 K for the $X_3$ transition in 6.9% MnP1. This value is about 3000 times larger than undoped $CsPbCl_3$ nanocrystals under identical conditions. The significant Zeeman splitting observed in these recently discovered perovskite QDs holds great potential for various applications in the future, such as spin-photonic quantum information processing, devices based on spin-filtering, magneto-optical gating, and charge-controlled magnetism.

**Author Contributions**

P.M. and R.V. formulated the project designs. P.M. conducted sample synthesis and performed all measurements. Data analysis and interpretation were reviewed by both P.M. and R.V. P.M. wrote the first draft, and both of them contributed to subsequent manuscript revisions.

**Notes**

The authors declare no competing financial interest.

**Acknowledgement**

The authors thank SERB (CRG/2018/000651), SERB-POWER Fellowship (SPF/2021/000110), ICMS, JNCASR, and the Department of Science and Technology, Government of India, Prof. C. N. R. Rao for financial support. P.M. thanks CSIR for a research fellowship. The authors acknowledge Technical Research Centre Microscopy lab at JNCASR Bangalore for TEM imaging.